# Machine learning formation enthalpies of intermetallics

Zhaohan Zhang[1], Mu Li[2], Katharine Flores[1,2] and Rohan Mishra[2,1,*]

[1]Institute of Materials Science & Engineering, Washington University in St. Louis, One Brookings Drive, St. Louis, MO 63130, USA

[2]Department of Mechanical Engineering & Materials Science, Washington University in St. Louis, One Brookings Drive, St. Louis, MO 63130, USA

*Corresponding author: rmishra@wustl.edu

**Abstract:**

Developing fast and accurate methods to discover intermetallic compounds is relevant for alloy design. While density-functional-theory (DFT)-based methods have accelerated design of binary and ternary alloys by providing rapid access to the energy and properties of the stable intermetallics, they are not amenable for rapidly screening the vast combinatorial space of multi-principal element alloys (MPEAs). Here, a machine-learning model is presented for predicting the formation enthalpy of binary intermetallics and used to identify new ones. The model uses easily accessible elemental properties as descriptors and has a mean absolute error (MAE) of 0.025 eV/atom in predicting the formation enthalpy of stable binary intermetallics reported in the Materials Project database. The model further predicts stable intermetallics to form in 112 binary alloy systems that do not have any stable intermetallics reported in the Materials Project database. DFT calculations confirm one such stable intermetallic identified by the model, $NbV_2$ to be on the convex hull. The model trained with binary intermetallics can also predict ternary intermetallics with similar accuracy as DFT, which suggests that it could be extended to identify compositionally complex intermetallics that may form in MPEAs.

**Introduction**

Intermetallics are metallic alloys that have a fixed composition and an ordered crystal structure. They form a diverse class of compounds with over 20,000 known compositions crystallizing in over 2,100 structure types with new ones being discovered constantly.[1] The presence of long range order and mixed bonding (metallic with ionic or covalent) in intermetallics distinguishes them from conventional metallic alloys in terms of their physical and mechanical properties. A common example is $Ni_3Al$ that crystalizes into an ordered $L1_2$ structure (according to the Pearson notation[2]) with Al atoms occupying the corners of a cube and Ni atoms occupying the face centers. As precipitates, it strengthens nickel-based superalloys for high-temperature applications.[3,4] In contrast, the formation of brittle intermetallics in the Au-Al systems, such as $AuAl_2$ (cF12-$CaF_2$ prototype), is a significant cause of wire bonding failures in microelectronics.[5,6] Besides their mechanical properties,[7,8] intermetallics are widely studied as shape memory alloys,[9] superconductors,[10] and catalysts.[11,12] With the rapid emergence of multi-principal element alloys (MPEAs), and especially high-entropy alloys (HEAs)[13] — that form a single-phase solid solution on mixing five or more elements at high (near equiatomic) concentration—, knowledge of the intermetallics that can form in such compositionally complex systems is vital to predict their microstructures and properties. For instance, the strengthening of ductile fcc $CoCrFeNiMo_x$ MPEAs has been attributed to the precipitation of hard $\sigma$ and $\mu$ intermetallic phases, without concomitant embrittlement.[14] Meanwhile, Troparevsky et al. have shown that the tendency of an MPEA to form a single-phase solid solution (HEA) can be predicted based on the formation enthalpy ($\Delta H_f$) of the pairwise binary intermetallics.[15] Therefore, fast and accurate prediction of intermetallics and their $\Delta H_f$ is of practical interest. But, it is challenging given the vast combinatorial space involving 81 elements and over 2,100 structure types.[1]

To efficiently navigate through this expansive chemical and structural space and rapdily screen intermetallic compounds, several strategies have been adapted. These range from empirical rules to high-throughput total energy calculations to machine learning (ML) models applied to materials databases.[16-19] Empirical, valence electron-counting rules have been successful in identifying elemental combinations that may form stable intermetallics within specific families such as Zintl and Heusler phases.[20, 21] With increasing computing power, first-principles density-functional-theory (DFT)-based high-throughput total energy calculations have allowed successful identification of new intermetallics without being confined to any particular structure type; however, they are inefficient to search for ternary and more chemically complex intermetallics where the configurational space and the computational expense explodes.[22-24] The availability of open-source materials databases, often built on results obtained from high-throughput DFT calculations, such as Materials Project (http://materialsproject.org/),[22] NOMAD (http://nomad-coe.eu/),[25] and OQMD (http://oqmd.org/),[23] have enabled the use of data-centric informatics methods — popularly called materials informatics — to identify new materials or predict unknown properties.[26] There are three ways to search for intermetallics using materials informatics: screening based on structure propotypes,[27] screening based on chemical compositions,[18] or a combination of the two.[19] While predicting new intermetallics with structure propotypes have efficiently reduced workload in interested systems,[19, 27] it is limited to common structure types. Here, our goal is to develop a machine learning model that can realize fast screening of intermetallics with good accuracy given just chemical composition.

In this Article, we present a ML model to accelerate the discovery of intermetallics by predicting their $\mathit{\Delta H_f}$ based on composition. The model was trained on DFT-calculated $\mathit{\Delta H_f}$ of stable intermetallics available in Materials Project, which lists the change in enthalpy upon forming an intermetallic with respect to the enthalpy of the constituent elements in their standard states.[22] We described metallic compositions using easily accessible elemental properties through the Matminer package,[28] and find properties like valence electron concentration, electronegativity, cohesive energy and first ionization energy to be the primary features of the model. The model can predict $\Delta H_f$ for binary intermetallics with a mean absolute error (MAE) of 0.022 eV/atom for the training set and an MAE of 0.044 meV/atom for the testing set. The model further predicts new, stable intermetallics to form in 112 metallic pairs where intermetallics have not been reported. We confirmed the stability of one of the predicted new binary intermetallics $NbV_2$ using DFT calculations and found it to exist as stable Laves phases. We applied the binary model to predict ternary intermetallics and found it reproduces the $\Delta H_f$ of known compounds with a mean absolute error (MAE) of 0.085 eV/atom and 0.057 eV/atom, without and with further training, respectively. Based on this result, we posit that this model can be extended to guide the prediction of compositionally complex intermetallics that may form in MPEAs.

**Methods**

We developed the model using the following three steps: building a dataset of binary intermetallics, describing the composition of each intermetallic with numerical attributes based on elemental properties, and mapping the attrubutes to $\mathit{\Delta H_f}$ through an ML algorithm. The workflow for the development of the model is shown schematically in Figure 1. We provide details about these three steps in the following.

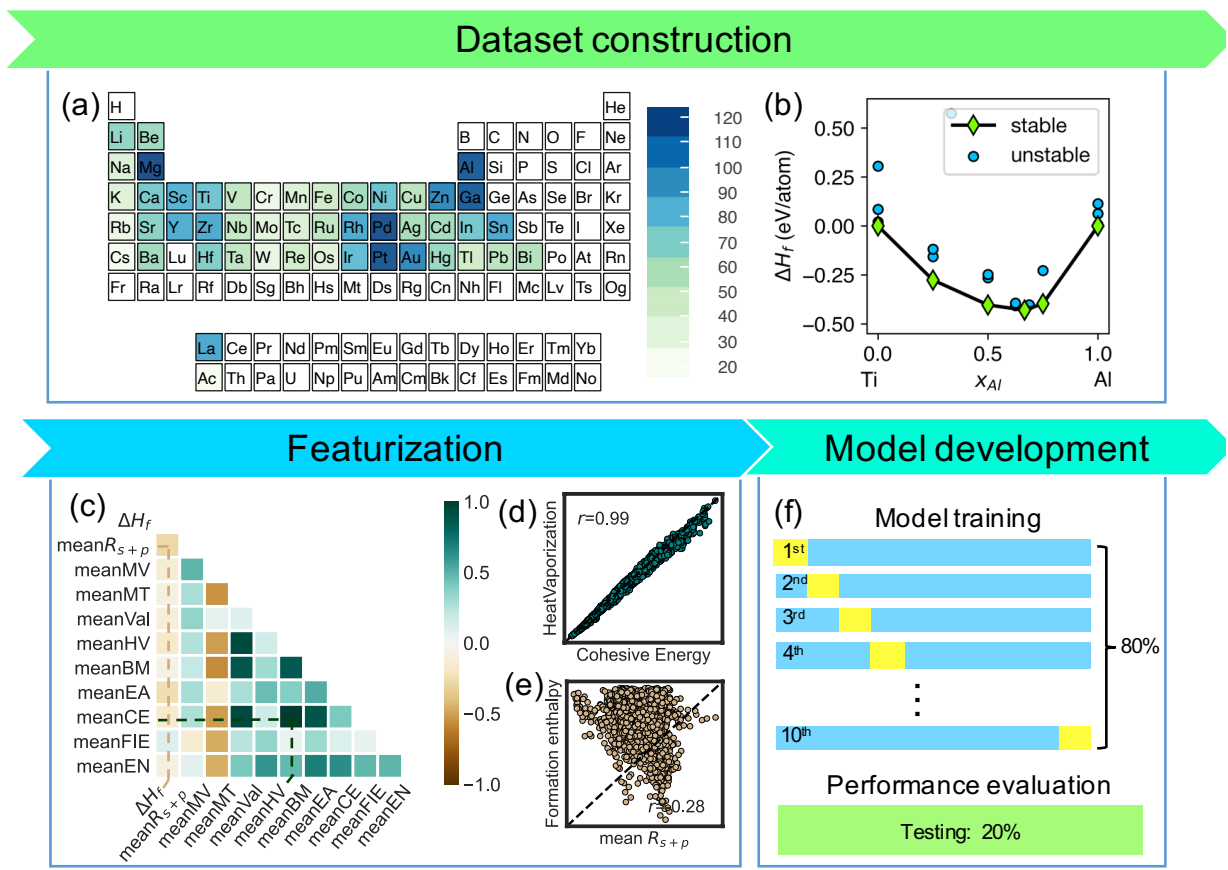

Fig. 1. Overall workflow for the statistical learning model. (a) Chemical space of intermetallics explored in the present study. Colormap indicates the frequency for each element to form stable binary intermetallics. (b) Binary phase diagram of Ti-Al. Stable intermetallics that are on the convex hull are highlighted with green color. (c) Pearson correlation matrix for elemental descriptors. Green and brown colors indicate positive and negative Pearson correlation coefficient, respectively. (d) Feature-feature scatter plot of cohesive energy and the heat of vaporization for binary intermetallics. (e) Feature-property scatter plot of Zunger's radius $R_{s+p}$ and formation enthalpy for binary intermetallics. (f) Schematic showing 10-fold cross-validation on 80% of the data used for model fitting. We use the remaining 20% to evaluate model performance.

*Datasets:* We picked binary intermetallics comprised of 48 metallic elements, including alkali metals, alkaline-earth metals, transition metals, post-transition metals, lanthanum, and actinium. These elements have been shaded with color in the Periodic Table shown in Fig. 1(a). We queried the Materials Project database for DFT-calculated $ΔH_f$ of all the stable binary intermetallics[22] using the Materials Application Programming Interface.[29] Figure 1(b) shows a typical binary phase diagram, in this case of Ti – Al, obtained from Materials Project. The solid black line indicates the convex hull that connects all the stable phases. Phases above the convex hull are metastable and are expected to decompose into adjacent stable phases under thermodynamic equillibrium. While training our model, we only include 1,538 stable binary intermetallics that are on the convex hull. The frequency with which the 48 elements form stable binary intermetallics is shown in Fig. 1(a) using a colormap. In a similar manner, we queried available phase diagrams of ternary systems in Materials Project and imported a list of 2,118 stable ternary intermetallics for evaluation.

*Compositional representation:* To develop an ML model, it is necessary to represent the various intermetallics numerically using one or more quantitative attributes, which are also called descriptors or features. The choice of descriptors is highly dependent on the property to be modeled. As we are interested in predicting the formability of intermetallics for any given combination of elements, we have used elemental properties to transform the chemical composition of intermetallics into numerical descriptors. We start by indexing several elemental properties, including Zunger's pseudopotential radius $R_{s+p}$ ($R_{s+p}$),[30] electronegativity (EN), molar volume (MV), melting temperature (MT), 1$^{st}$ ionization energy (FIE), number of valence electrons (Val), the heat of vaporization (HV), electron affinity (EA) as saved in the Magpie preset in Matminer,[28,31] and bulk modulus (BM) and cohesive energy (CE) collected from the literature.[32] We used the Matminer package to transform the elemental properties to compositional descriptors. For every alloy composition, we map each elemental property using 4 attributes: the mean, mean absolute deviation, maximum and minimum value of the composition's consitituent elements, which results in a total of 40 descriptors. To avoid linearly correlated descriptors, we calculated the Pearson correlation between each pair of descriptors. A correlation map for the mean attributes of the elemental properties are shown in Fig. 1(c). The colormap indicates the Pearson correlation coefficient. This coefficient ranges from −1 to 1, with −1 denoting total negative linear correlation, 0 representing no linear correlation, and 1 showing total positive linear correlation. For those elemental properties that are very strongly correlated (Pearson correlation coefficient larger than 0.9),[33] one set is eliminated based on domain knowledge. For example, cohesive energies of elements show a strong linear relatonship with their heats of vaporization, as shown in Fig. 1(d). We retain only cohesive energy as a feature since it represents the energy gained by arranging atoms to a crystalline state, while heat of vaporization is the energy needed to transfer a liquid to its gaseous state. After removing highly correlated elemental properties, we are left with 32 descriptors. None of these elemental properties show a strong linear correlation with $\Delta H_f$; as an example, we show the variation of $\Delta H_f$ with the mean pseudopotential radius $R_{s+p}$ in Fig. 1(e). Therefore, we use nonlinear regression functions as discussed below.

*Statistical Learning Methods:* We employed gaussian process regression (GPR) as implemented in the Scikit-learn python package to learn the nonlinear relationship between an intermetallic's descriptors and its $\Delta H_f$.[34] For a given set of numerical descriptors, GPR learns a multivariate gaussian distribution that maps them to the target property $(\Delta H_f)$. We use a sum-kernel which consists of a squared-exponential kernel and a white kernel that represents noise:

$$k(x_i, x_j) = \sigma^2 exp\left(-(x_i - x_j)^2/2l^2\right) + \sigma_n^2 \delta(x_i, x_j), \tag{1}$$

where $x_i$ and $x_j$ are descriptor vectors for two intermetallics $i$ and $j$; signal variance $\sigma$, length-scale parameter $l$, and noise level $\sigma_n$ are hyperparameters to be determined during the training process. We also tested other nonlinear regression methods with the same kernel, such as kernel ridge regression.[35] We chose GPR because it learns a generative, probabilistic model of the target function and can thus provide meaningful uncertainties/confidence intervals along with the predictions.[36] Also, GPR can choose the kernel's hyperparameters based on gradient-ascent on the marginal likelihood function, which is relatively fast compared to similar models.

Before training the model, all descriptors are rescaled to range [0, 1] with min-max normalization. Here elements $x_i$ of a descriptor vector $x = (x_1, x_2, …, x_n)$ is normalized to $x_i'$ with $x_i' = \frac{x_i - min(x)}{max(x) - min(x)}$. We used 48 metallic elements and 80% of the binary dataset for model training (training set), and kept the remaining 20% for model evaluation (testing set). To generalize the model and estimate model performance on new data, we applied 10-fold cross validation (CV) during the training process, which randomly splits the data into 10 subsets, and iteratively fits the model with nine of them and evaluates on the remaining subset. The accuracies are averaged for the entire process and defined as CV accuracy. This process is shown schematically in Fig. 1(f).

*Computational Methods:* We performed the DFT calculations using the Vienna Ab-initio Simulation Package (VASP) with projector augmented-wave potentials.[37, 38] For the search of stable intermetallic structures, we employed generalized gradient approximation (GGA) as implemented in the Perdew−Burke−Ernzerhof (PBE) functional.[39] For $\Delta H_f$ calculations, we used a plane-wave basis set with a cutoff energy of 400 eV and performed relaxation until the Hellmann-Feynman forces on the atoms are less than 0.001 eV/Å. The Brillouin zone was sampled using a Monkhorst−Pack *k*-points mesh while keeping the number of *k*-points times lattice constant equal to ~30 and ~80 for structural relaxation and the single-step static calculation, respectively.[40] The phonon calculations were performed using the frozen-phonon approach, and the dispersion spectra were calculated using the Phonopy package.[41] For accurate phonon calculations, a higher cutoff energy of 700 eV for the plane-wave basis set was used with a tighter electronic convergence of $10^{-8}$ eV. Additionally, to calculate the force-constant matrices, a $2 \times 2 \times 2$ supercell was used for the C15 phase and a $3 \times 3 \times 2$ supercell was used for the C14 phase.

**Results and Discussion**

To identify the most important set of descriptors that can accurately predict $\Delta H_f$, we employed forward and backward feature selection during the model development. As mentioned above, we use 4 attributes (mean, mean abosolute deviation, maximum and minimum) of each elemental property as descriptors to distinguish between different compositions of any pair of elements. In forward selection, we start with an empty feature set and iterate through the various features and select one that maximizes the CV accuracy. We found that the VEC, i.e., the mean, mean absolute deviation, maximum and minimum VEC values, leads to the highest CV accuracy of 0.64, as measured by the coefficient of determination $R^2$. We then kept these 4 attributes in the feature set and searched for a second set of attributes of an elemental property that maximizes the CV accuracy. We cycled this process until the addition of a new property did not further improve the performance of the model. This process is shown in Fig. 2(a) with the colormap indicating CV accuracy and yellow stars indicating the elemental property that is selected in each iteration. Using this approach, we reduced the total number of descriptors from 32 to 16, with the best subset including VEC, FIE, EN, and CE. The elemental properties are indexed in the table in Fig. 2(a).

The improvement in the accuracy of the GPR model with an increasing number of descriptors (also the number of iterations) during the forward feature selection process is shown in Fig. 2(b). The coefficient of determination $R^2$ for the 10-fold CV (average accuracy for the 10 CV iterations) prediction on training and testing data increases until the model reaches 16 decriptors, after which it either plateaus or decreases slightly due to overfitting.

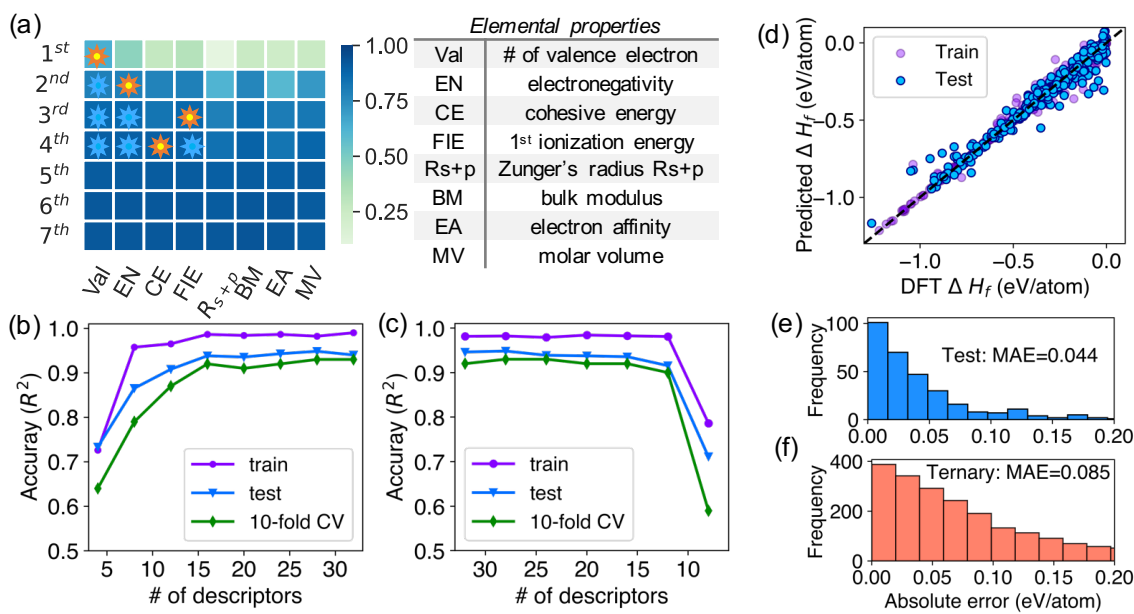

Fig. 2. Feature selection and model evaluation for binary intermetallics. (a) Schematic showing the process of forward feature selection. Elemental property descriptors are indexed as shown in the table. The variations of training, testing, and cross validation accuracies (in terms of coefficient of determination $R^2$) during (b) forward and (c) backward feature selection. (d) Performance of a 12-feature model for binary intermetallics. Model predicted formation enthalpies are plotted with respect to DFT-calculated values; Absolute errors for the testing set are shown separately in (e). Using this model trained with binary intermetallics to make prediction for ternary intermetallics, absolute errors for ternary intermetallics are shown in (f).

To further validate the importance of the descriptors down-selected from forward selection and ascertain that important descriptors were not left out, we used backward eliminaton. Here, we start with all the 32 descriptors and iteratively drop the one that has the least effect on accuracy, until a significant drop is observed on further removal of any descriptor. The variation of accuracies during this process is shown in Fig. 2(c), which also reduced to 16 descriptors with the same subset as forward selection.

In the above selection process, we simultaneously added or eliminated all 4 attributes of any elemental property. To further reduce the risk of overfitting, we then performed feature selection with respect to individual attributes of the down-selected elemental properties. We found that the mean absolute deviation attributes for elemental properties did not play a role in improving the model performance, which reduces the number of descriptors to 12. We obtained a mean absolute error (MAE) of 0.022 eV/atom for the training set and a MAE of 0.044 eV/atom for the testing set with this 12-feature model, as shown in Fig. 2(d). For reference, the MAE of DFT calculated $\Delta H_f$ with respect to experimental measurements is ~ 0.145 eV/atom for entries in the Materials Project database, when using the elemental DFT total energies as chemical potentials.[42,43] Furthermore, 80% of the binary intermetallics are predicted within an absolute error of 0.025 eV/atom using our model for the testing set, as shown in Fig. 2(e). These results indicate that the model predictions are reliable with DFT-level accuracy and it can potentially be applied for the discovery of new intermetallic compositions.

To further establish the generality of our model to predict $\Delta H_f$ of complex intermetallics, we evaluate the model that was trained with binary intermetallics with a list of 2,118 stable ternary intermetallics obtained from the Materials Project. The 12-feature model, without any further training, gives a MAE of 0.085 eV/atom in the predicted $\Delta H_f$ of the ternary intermetallics. The inset in Fig 2(f) shows the histogram of the absolute error compared to DFT-calculated $\Delta H_f$ for ternary intermetallics. During the feature selection process, we noticed that although the CV accuracy for the model to predict $\Delta H_f$ of binary intermetallics saturates with ~ 12 descriptors, its prediction accuracy in the case of ternary intermetallics increases with more descriptors, as shown in the forward feature selection curve in Fig. 3(a). By increasing the number of descriptors from 12 to 24, MAE in $\Delta H_f$ of ternary intermetallics decreases from 0.085 eV/atom to 0.057 eV/atom.

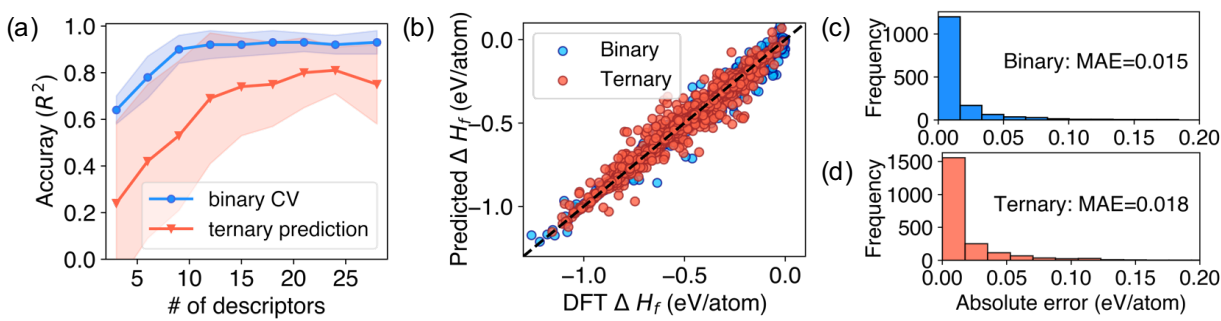

Fig. 3: Applying binary model to ternary intermetallics. (a) Forward feature selection indicates that increasing descriptor number gives better accuracy on ternary intermetallics. Shaded curves represent the mean CV accuracies and their standard derivation. (b) A 24-feature model for both binary and ternary intermetallics. Model predicted formation enthalpies are plotted with respect to DFT-calculated values; Absolute errors for the binary intermetallics and ternary intermetallics are shown in (c) and (d), respectively.

We then combined the dataset of ternary intermetallics with that of the binary intermetallics and develop a 24-feature model that was trained with 80% of the entries in the combined dataset, and evaluated with the remaining 20%. We found that the model predicted $\Delta H_f$ values agree well with those calculated using DFT for both binary and ternary intermetallics, which are plotted seperately in Fig. 3(b). This observed improvement in accuracy is due to the more than two-fold increase in the amount of training data. The histograms of absolute error for binary and ternary intermetallics are shown in Fig. 3(c) and Fig. 3(d), respectively. We propose that by similar conditioning to a limited set of multielement intermetallic systems, the present model could be extended to predict the formability of intermetallics from a compositionally vast space involving multiple elements (4, 5 or even more), whose $\Delta H_f$ are rarely available from first-principles-based databases. The performance of our model in terms of MAE in the predicted $\Delta H_f$ is comparable to other recently proposed models: ElemNet, whose training set size was 230,960 compounds has a MAE of 0.055 eV/atom on the testing set;[44] CGCNN, which consists both compositional and structural descriptors with 28,046 training data, has a MAE of 0.039 eV/atom on the testing set.[45]

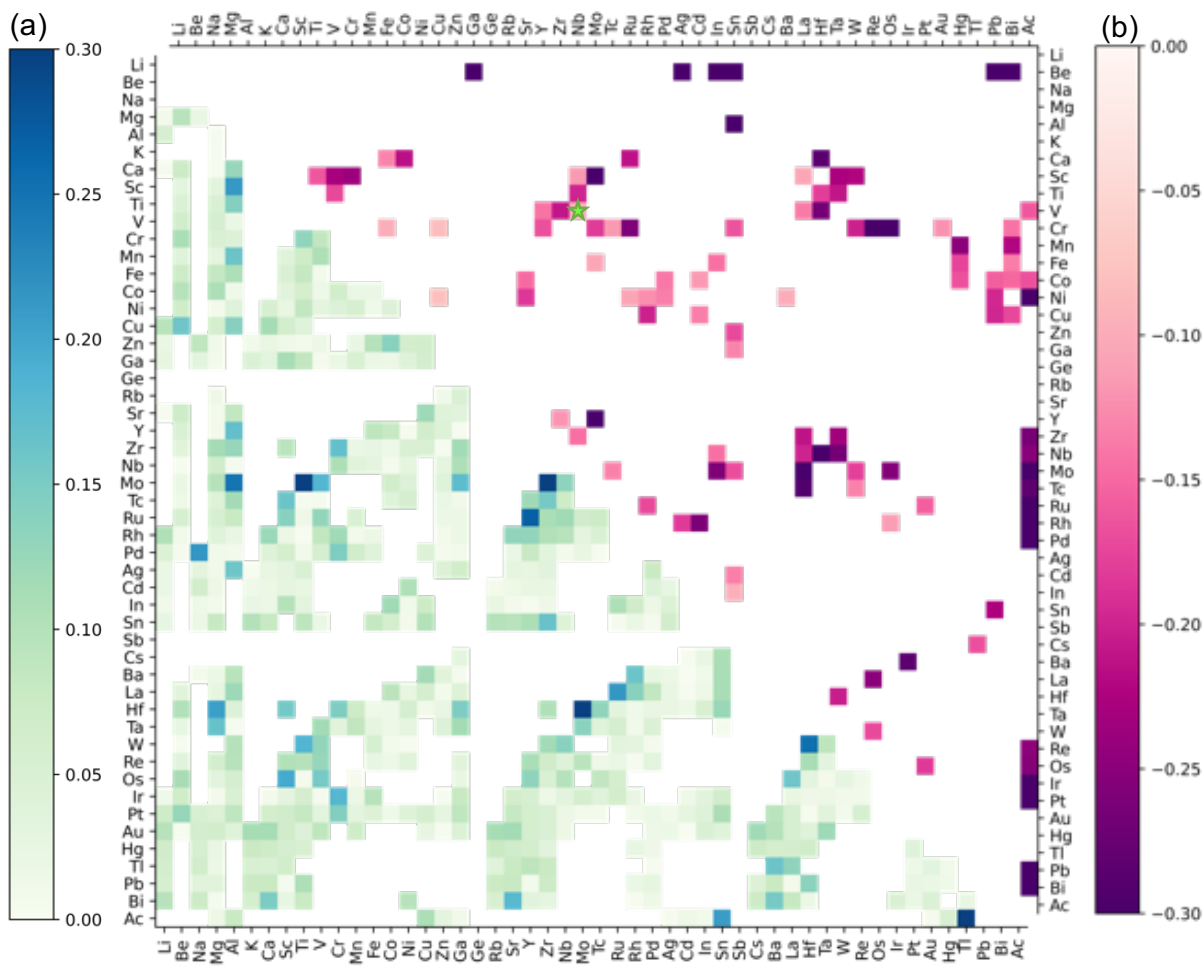

Fig. 4. Binary intermetallic screening map. The lower triangle shows the absolute prediction error on existing binary intermetallic systems. The colormap indicates the absolute error in units of eV/atom. The upper triangle shows new binary intermetallic systems that are predicted with a high probability. The colormap indicates the predicted lowest formation enthalpy value in units of eV/atom.

Having established the capability of the model to accurately capture $\Delta H_f$ of known binary and ternary intermetallics, we used it to explore the possibility of finding new binary intermetallics that have not been reported. We screened all binary combinatorial compositions of the 48 metallic elements, which results in 1,128 unique pairs with 603 of them reported in Materials Project. For each pair of elements, whether stable intermetallic compounds have been reported or not, a list of binary compositions ranging from $A_{0.1}B_{0.9}$ to $A_{0.9}B_{0.1}$ (10% mole concentration intervals) is transferred into numerical attributes. We then use the developed model to predict their $\Delta H_f$ and construct a convex hull. The lowest $\Delta H_f$ on the convex hull is recorded. For those binary intermetallic systems that have been reported, the predicted convex hull agrees well with DFT results from the Materials Project. As an example, the convex hull of Ti-Al predicted by the model is compared to that obtained from Materials Project in Appendix Fig. 6. The absolute error of the predicted lowest $\Delta H_f$ with respect to the DFT-calculated value for the reported 603 binary pairs can be visualized in the colormap in Fig. 4(a). 80% of the known binary intermetallics are predicted within an error of 0.025 eV/atom. Furthermore, we predict 112 new metallic pairs to form stable intermetallics with high probability, as shown in Fig. 4(b). By high probability, we mean those intermetallics whose bounds for the predicted distribution of $\Delta H_f$ are negative (since GPR can make probabilistic prediction with meaningful uncertainty). The colormap in Fig. 4(b) indicates the mean prediction value of the lowest $\Delta H_f$ for each pair of elements. This makes up 21% of pair grids that were blank in Materials Project, which indicates that there may be more stable binary intermetallic compounds waiting to be explored. We have included a more exhaustive map of ML-predicted combinatorial screening formability in the appendix Fig. 7. As suggested by Troparevsky *et al.*, the lowest $\Delta H_f$ for each pair of elements can be used as an important tool for assessing the formation of single-phase HEAs.[15]

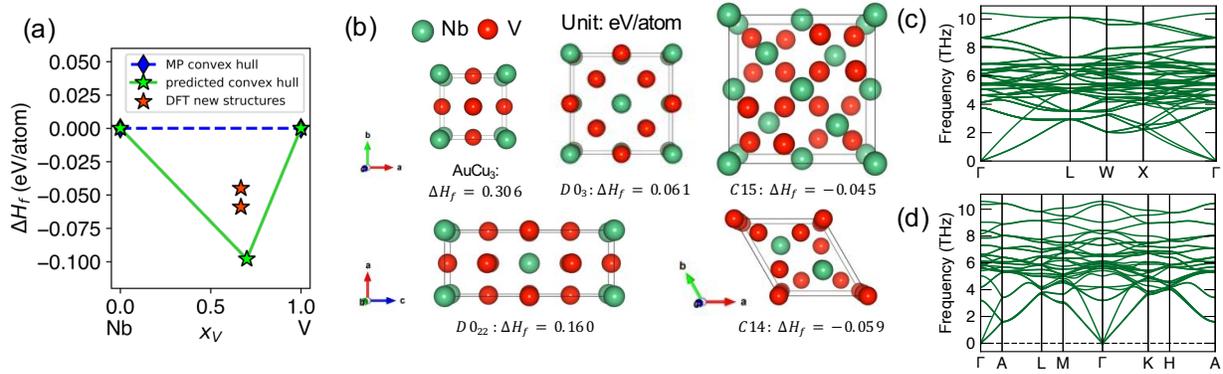

Fig. 5. (a) Model predicted Nb-V binary phase diagram: a new intermetallic is predicted with a composition near $Nb_{0.7}V_{0.3}$. (b) DFT calculations for some intermetallic structures: two Laves structures (C15 and C14) with composition $NbV_2$ are confirmed to have a negative formation enthalpy. Their phonon band structures are plotted in (c) and (d), respectively.

One of the binary combinations that does not have any stable binary intermetallics reported in Materials Project is Nb-V, which is shown as a horizontal line (blue) in the phase diagram in Fig. 5(a). The combination is labeled in Fig. 4(b) with a green star. Nb and V alloys have an miscibility gap below 500 K.[46] However, our model predicts a stable intermetallic for $Nb_{0.3}V_{0.7}$ with a $\Delta H_f$ of −0.097 eV/atom and a credible interval 0.060 eV/atom, which suggests with 95% probability that its $\Delta H_f$ should range between −0.157 and −0.037 eV/atom. To confirm this prediction, we performed DFT calculations. As our model does not capture crystal structures, we screened structure prototypes that are formed by refractory metals (Nb, Zr, Ti, V) near this composition on the Inorganic Crystal Structure Database (ICSD) and the Materials Project.[47] The structures and compositions we found in ICSD are listed in Table 1 in the appendix. We then calculated their $\Delta H_f$ using DFT by substituting Nb and V into these structures; some of these are shown in Fig. 5(b). We found two Laves structures, C15 and C14 to be thermodynamically stable, with $\Delta H_f$ values of −45 eV/atom and −59 eV/atom, respectively, calculated with respect to bcc V and hcp Nb. In both of the structures, V atoms form tetrahedra around Nb with Nb atoms ordered either in a diamond cubic structure (C15-Laves) or in hexagonal structure (C14-Laves). To further confirm the dynamical stability of these Laves phases, we performed phonon calculations. We did not observe any soft modes in their phonon band structure plots shown in Fig. 5(c).

The combination of our ML model with DFT calculations can accelerate the discovery of new, compositionally complex intermetallics. As a general strategy, we propose that for a given combination of elements, our model be used to rapidly identify compositions that are predicted to form stable intermetallics with high probability. One can then substitute the elements in the identified intermetallic compositions with chemically similar elements, for instance by using the probabilistic model developed by Hautier et al.,[48, 49] to obtain a set of similar intermetallics, with the expectation that some of them may have been reported previously. This should be followed by a search in existing databases to screen the various crystal structures adopted by the identified intermetallics. Subsequently, DFT calculations should be used to optimize the targeted composition with the screened structures and determine their thermodynamic stability.

**Conclusions**

We have developed a fast and accurate ML model to predict the formability of binary/ternary intermetallics given any metallic pair/triplet combination. The model achieves strong performance within the range of metallic alloys. Our 12-feature model for binary intermetallics predicts the formation enthalpy of the testing set with a MAE of 0.044 eV/atom.

Our model enables the screening of millions of compositions within seconds, which is ideal for exploring the vast combinatorial space for compositionally complex intermetallics. Meanwhile, we can use the current model developed for binary and ternary intermetallics to predict the stability of intermetallic phases in MPEAs. A statistical study on 142 intermetallic-containing MPEAs has shown that all of the intermetallic phases contained in the MPEAs are existing structures in the binary/ternary subsystems of their respective alloys.[50] Therefore, by predicting $\Delta H_f$ for a list of binary/ternary compositions formed by each pair/triplet of the constituent elements with our model, and quantitatively comparing with configurational entropy of the multi-element system at different temperatures, we can predict whether the stable phase of MPEAs would be an ordered structure, disordered solid solution or a combination of both. Instead of directly searching the combinatorial space of 5 or more elements, our model provides another prospective method to predict the intermetallic phases likely to form in MPEAs.

To facilitate the discovery of new intermetallics, we are making our model available at: https://github.com/M-cube-wustl/ML_intermetallics

**Acknowledgments:** This work was supported by the National Science Foundation (NSF) through grant number DMR-1809571. This work used the computational resources of the Extreme Science and Engineering Discovery Environment (XSEDE), which is supported by NSF ACI-1548562. Authors are grateful to Prof. Rampi Ramprasad and Dr. Logan Ward for fruitful discussions.

**Data availability:** Data available in article or supplementary material.

**Appendix:**

Table 1. Structures used for searching the crystal structure of predicted Nb-V intermetallic.

| Composition | ICSD-ID | MP-ID | Structure type | Space group |
|---|---|---|---|---|
| (Nb, Ti) | 105248 | NaN | bcc-W | Im-3m (229) |
| $Ti_3Nb$ | 671498 | mp-980945 | Auricupride-$Cu_3Au$ | Pm3m (221) |
| $Ti_3Nb$ | NaN | mp-1187514 | hexagonal | $P6_3$/mmc (194) |
| $TiAl_3$ | 58189 | NaN | $TiAl_3$ | I4/mmm (139) |
| $V_2Zr$ | 106214 | mp-258 | Laves(cub)-$MgCu_2$ | Fd-3m (227) |
| $V_2Zr$ | 653414 | NaN | Laves(2H)-$MgZn_2$ | $P6_3$/mmc (194) |
| $ZrTi_2$ | 247962 | mp-1008568 | $CaHg_2$ | P6/mmm (191) |

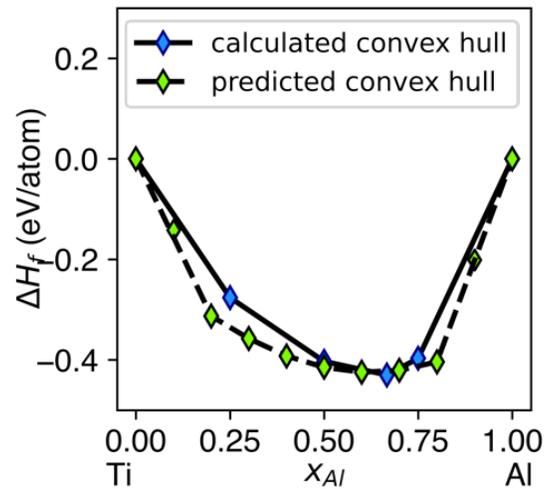

Fig. 6. Model-predicted convex-hull diagram for Ti-Al is shown in dash line with green labels indicating the prediction value range from $Ti_{0.1}Al_{0.9}$ to $Ti_{0.9}Al_{0.1}$. For reference, DFT calculated convex hull from Materials Project is shown in solid line with blue labels.

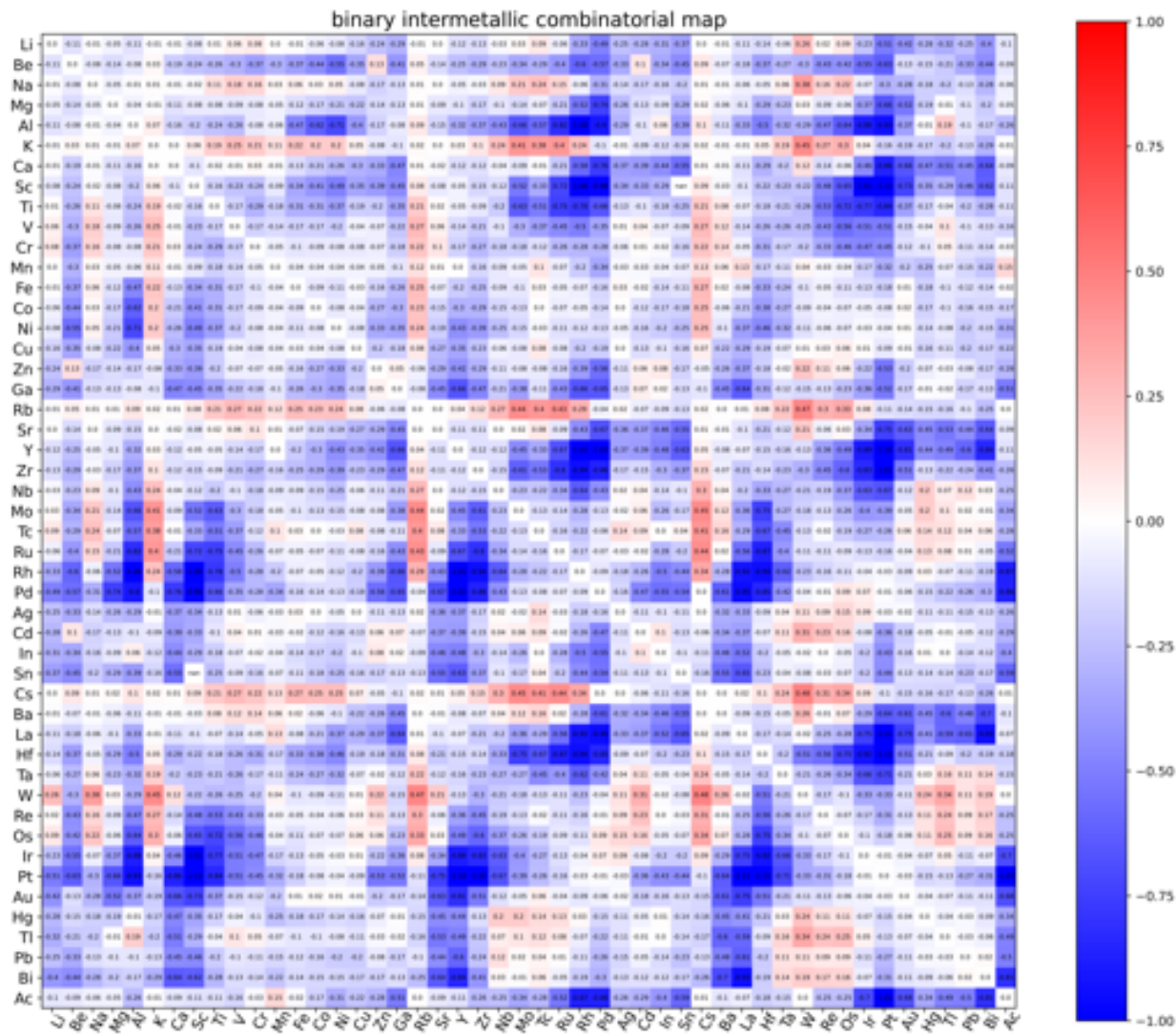

Fig. 7. Model-predicted lowest formation enthalpy for each pair of elements. The colormap shows the formation enthalpy in units of eV/atom; the actual value is also annotated.

**References:**

1. J. Dshemuchadse and W. Steurer, Inorg Chem **54** (3), 1120-1128 (2015).
2. W. B. Pearson, *A Handbook of Lattice Spacings and Structures of Metals and Alloys*. (Elsevier, 1958).
3. L. Kovarik, R. R. Unocic, J. Li, P. Sarosi, C. Shen, Y. Wang and M. J. Mills, Progress in Materials Science **54** (6), 839-873 (2009).
4. R. W. Kozar, A. Suzuki, W. W. Milligan, J. J. Schirra, M. F. Savage and T. M. Pollock, Metallurgical and Materials Transactions A **40** (7), 1588-1603 (2009).
5. G. G. Harman and G. G. Harman, *Wire bonding in microelectronics : materials, processes, reliability, and yield*. (McGraw-Hill, New York, 1997).
6. H. Xu, C. Liu, V. V. Silberschmidt, S. S. Pramana, T. J. White, Z. Chen and V. L. Acoff, Intermetallics **19** (12), 1808-1816 (2011).
7. A. I. Taub and R. L. Fleischer, Science **243** (4891), 616-621 (1989).
8. P. Jozwik, W. Polkowski and Z. Bojar, Materials **8** (5), 2537-2568 (2015).
9. K. Otsuka and X. B. Ren, Intermetallics **7** (5), 511-528 (1999).
10. R. J. Cava, H. Takagi, H. W. Zandbergen, J. J. Krajewski, W. F. Peck, T. Siegrist, B. Batlogg, R. B. Vandover, R. J. Felder, K. Mizuhashi, J. O. Lee, H. Eisaki and S. Uchida, Nature **367** (6460), 252-253 (1994).
11. A. Dasgupta and R. M. Rioux, Catalysis Today **330**, 2-15 (2019).
12. D. Sun, Y. Wang, K. J. T. Livi, C. Wang, R. Luo, Z. Zhang, H. Alghamdi, C. Li, F. An, B. Gaskey, T. Mueller and A. S. Hall, ACS Nano **13** (9), 10818-10825 (2019).
13. For a discussion on the taxonomy of MPEAs and HEAs, see the review by Miracle and Senkov. D. B. Miracle and O. N. Senkov, Acta Materialia 122, 448-511 (2017).
14. W. H. Liu, Z. P. Lu, J. Y. He, J. H. Luan, Z. J. Wang, B. Liu, Y. Liu, M. W. Chen and C. T. Liu, Acta Mater. **116**, 332-342 (2016).
15. M. C. Troparevsky, J. R. Morris, P. R. C. Kent, A. R. Lupini and G. M. Stocks, Phys. Rev. X **5** (1), 6 (2015).
16. A. R. Miedema, F. R. d. Boer and P. F. d. Chatel, Journal of Physics F: Metal Physics **3** (8), 1558-1576 (1973).
17. K. Kim, L. Ward, J. He, A. Krishna, A. Agrawal and C. Wolverton, Physical Review Materials **2** (12), 123801 (2018).

18. S. Ubaru, A. Międlar, Y. Saad and J. R. Chelikowsky, Physical Review B **95** (21), 214102 (2017).
19. J. Schmidt, L. M. Chen, S. Botti and M. A. L. Marques, J. Chem. Phys. **148** (24), 6 (2018).
20. H. Schäfer, B. Eisenmann and W. Müller, Angewandte Chemie International Edition in English **12** (9), 694-712 (1973).
21. J. He, S. S. Naghavi, V. I. Hegde, M. Amsler and C. Wolverton, Chemistry of Materials **30** (15), 4978-4985 (2018).
22. A. Jain, S. P. Ong, G. Hautier, W. Chen, W. D. Richards, S. Dacek, S. Cholia, D. Gunter, D. Skinner, G. Ceder and K. A. Persson, APL Mater. **1** (1), 11 (2013).
23. S. Kirklin, J. E. Saal, B. Meredig, A. Thompson, J. W. Doak, M. Aykol, S. Ruhl and C. Wolverton, npj Comput. Mater. **1**, 15 (2015).
24. S. Curtarolo, W. Setyawan, G. L. W. Hart, M. Jahnatek, R. V. Chepulskii, R. H. Taylor, S. D. Wanga, J. K. Xue, K. S. Yang, O. Levy, M. J. Mehl, H. T. Stokes, D. O. Demchenko and D. Morgan, Comp Mater Sci **58**, 218-226 (2012).
25. L. M. Ghiringhelli, C. Carbogno, S. Levchenko, F. Mohamed, G. Huhs, M. Lüders, M. Oliveira and M. Scheffler, npj Comput. Mater. **3** (1), 46 (2017).
26. R. Ramprasad, R. Batra, G. Pilania, A. Mannodi-Kanakkithodi and C. Kim, npj Computational Materials **3** (1) (2017).
27. A. O. Oliynyk and A. Mar, Accounts of Chemical Research **51** (1), 59-68 (2018).
28. L. Ward, A. Dunn, A. Faghaninia, N. E. R. Zimmermann, S. Bajaj, Q. Wang, J. Montoya, J. Chen, K. Bystrom, M. Dylla, K. Chard, M. Asta, K. A. Persson, G. J. Snyder, I. Foster and A. Jain, Comp Mater Sci **152**, 60-69 (2018).
29. S. P. Ong, S. Cholia, A. Jain, M. Brafman, D. Gunter, G. Ceder and K. A. Persson, Comput. Mater. Sci. **97**, 209-215 (2015).
30. A. Zunger, Physical Review B **22** (12), 5839-5872 (1980).
31. L. Ward, A. Agrawal, A. Choudhary and C. Wolverton, npj Comput. Mater. **2** (1), 16028 (2016).
32. C. Kittel, P. McEuen, W. John and Sons, *Introduction to solid state physics*. (John Wiley & Sons, Hoboken, NJ, 2019).
33. P. Schober, C. Boer and L. A. Schwarte, Anesthesia & Analgesia **126** (5) (2018).

34. L. Buitinck, G. Louppe, M. Blondel, F. Pedregosa, A. Mueller, O. Grisel, V. Niculae, P. Prettenhofer, A. Gramfort, J. Grobler, R. Layton, J. Vanderplas, A. Joly, B. Holt and G. Varoquaux, API Design for Machine Learning Software: Experiences from the Scikit-learn Project (2013).

35. K. P. Murphy, *Machine Learning: A Probabilistic Perspective*. (The MIT Press, 2012).

36. C. E. Rasmussen and C. K. I. Williams, *Gaussian Processes for Machine Learning (Adaptive Computation and Machine Learning)*. (The MIT Press, 2005).

37. G. Kresse and J. Furthmuller, Physical Review B **54** (16), 11169-11186 (1996).

38. P. E. Blöchl, Physical Review B **50** (24), 17953-17979 (1994).

39. J. P. Perdew, K. Burke and M. Ernzerhof, Physical Review Letters **77** (18), 3865-3868 (1996).

40. H. J. Monkhorst and J. D. Pack, Physical Review B **13** (12), 5188-5192 (1976).

41. A. Togo and I. Tanaka, Scripta Materialia **108**, 1-5 (2015).

42. A. Jain, G. Hautier, C. J. Moore, S. Ping Ong, C. C. Fischer, T. Mueller, K. A. Persson and G. Ceder, Comput. Mater. Sci. **50** (8), 2295-2310 (2011).

43. A. Jain, G. Hautier, S. P. Ong, C. J. Moore, C. C. Fischer, K. A. Persson and G. Ceder, Physical Review B **84** (4), 045115 (2011).

44. D. Jha, L. Ward, A. Paul, W.-k. Liao, A. Choudhary, C. Wolverton and A. Agrawal, Scientific Reports **8** (1), 17593 (2018).

45. T. Xie and J. C. Grossman, Physical Review Letters **120** (14), 145301 (2018).

46. metallab.net/chemsoc/alloys.php?elementno=41&index=2 (Accessed on 4/20/2020).

47. M. Hellenbrandt, Crystallography Reviews **10** (1), 17-22 (2004).

48. G. Hautier, C. C. Fischer, A. Jain, T. Mueller and G. Ceder, Chemistry of Materials **22** (12), 3762-3767 (2010).

49. G. Hautier, C. Fischer, V. Ehrlacher, A. Jain and G. Ceder, Inorg Chem **50** (2), 656-663 (2011).

50. R.-C. T. Ming-Huang Tsai, Ting Chang, Wei-Fei Huang, Metals (2019).